\documentstyle[12pt]{article}
\def\ref#1{\noindent\hangindent=24.0pt\hangafter=1{#1}\par}
\def\la{\hbox{\rlap{$<$}\lower.5ex\hbox{$\sim$}\ }}
\def\ga{\hbox{\rlap{$>$}\lower.5ex\hbox{$\sim$}\ }}
\begin{document}
\begin{center}

\bigskip

		{\bf X-ray variability in M87} \\

\bigskip

	D. E. Harris \\
	SAO MS-3, 60 Garden St., Cambridge MA 02138 USA\\
	harris@cfa.harvard.edu\\

	J. A. Biretta\\
	 STScI, 3700 San Martin Dr., Baltimore MD 21218 USA\\
	biretta@stsci.edu \\

	W. Junor\\
	The Institute for Astrophysics,
        Dept. of Physics and Astronomy, Univ. of New Mexico,\\
	800 Yale Blvd., NE Albuquerque, NM 87131 USA\\
	bjunor@astro.phys.unm.edu\\

\bigskip

Received 1996 
 
\end{center}
\bigskip
\section*{ABSTRACT}
	
	We present the evidence for X-ray variability from the core
and from knot A in the M87 jet based on data from two observations
with the Einstein Observatory High Resolution Imager (HRI) and three
observations with the ROSAT HRI.  The core intensity showed a 16\%
increase in 17 months ('79-'80); a 12\% increase in the 3 years '92 to
'95; and a 17\% drop in the last half of 1995.  The intensity of knot
A appears to have decreased by 16\% between 92Jun and 95Dec.  Although
the core variability is consistent with general expectations for AGN
nuclei, the changes in knot A provide constraints on the x-ray
emission process and geometry.  Thus we predict that the x-ray
morphology of knot A will differ significantly from the radio and
optical structure.

\bigskip

Key words: galaxies:active - galaxies:individual:M87 - galaxies:jet -
X-rays:galaxies

\newpage
\setcounter{section}{0}
\section{INTRODUCTION}

	The Einstein Observatory (EO) HRI observations of M87 were the
first to clearly isolate X-ray emission of the core of the galaxy and
of the brightest knot in the jet from the broader distributions known
previously (Schreier, Gorenstein \& Feigelson 1982).  These authors
suggested that the core emission was resolved, and thus likely to be
thermal bremsstrahlung whereas the emission from knot A was probably
synchrotron emission.  The EO data were further analyzed by Biretta,
Stern \& Harris 1991 (hereafter `BSH'), who summed the two EO/HRI
observations and selected only a portion of the data in order to
achieve the best possible image integrity.  BSH argued that most of
the core emission was unresolved and thus could be similar to nuclear
emission from other AGN.  Since AGN exhibit X-ray variability with
timescales of days to years (e.g. Mushotzky, Done \& Pounds 1993),
the same behavior could occur in M87.

	When the ROSAT archival data became publicly available, it was
evident that the intensity ratio of the core to knot A had changed,
and we thus proposed further observations with the ROSAT HRI.  In this
paper we report only on the gross intensity changes since poor aspect
solutions (which have an effect similar to `pointing jitter') have
degraded the effective resolution of the 1995 data.  Consequently, the
sizes of the regions used to measure fluxes contain a higher
percentage of background emission than desirable.  Despite these
difficulties, we find convincing evidence for variability in both the
core (+ knot D) and in knot A (+ knot B).

	The most reliable evidence for variability comes from a
comparison between the multiple observations made with each satellite.
When comparing Einstein and ROSAT data however, the difference in
effective area as a function of energy between the two satellites
introduces an uncertainty because we have no knowledge of the X-ray
spectral distributions of the various components.

\section{DATA REDUCTION}

\subsection{The data}

The observations used in this paper are summarized in Table 1.  The
ROSAT images are shown in Figure 1, and may be compared to Figures 1
and 2 of BSH.  It is apparent that the 1995 data suffer from severe
aspect smearing and that the intensity ratio of the core to knot A has
increased.  The degradation in resolution from the aspect problem is
not easily fixed.  The only other source in the field with a
reasonable intensity is too far off axis (12.2$^{\prime}$) to serve as
a template for a point response function.

\subsection{Selection of regions for measuring the intensities}

To derive reliable intensities, we need to ensure that we collect the
same fraction of source counts for each feature and each observation.
Poor aspect degrades the resolution.  N/S profiles on the maps
smoothed with a 3$^{\prime\prime}$ Gaussian give FWHMs of
7.5$^{\prime\prime}$ for the core and 7.2$^{\prime\prime}$ for knot A
in the 92Jun data, but 10.0$^{\prime\prime}$ for both features in the
95Jun data.  Therefore, small radii circles would not measure the same
fraction of counts for different observations.  Although larger
integration areas can overcome the aspect smearing, they will suffer
from greater contamination with (non-variable) background emission.
Moreover, the core and knot A are separated by only
12$^{\prime\prime}$.  Consequently, we have made two sets of intensity
measurements: one by selecting adjoining boxes, and the other with
circular apertures of radius, r=6$^{\prime\prime}$.  The former will
be used for comparing countrates but will underestimate any
variability because of contamination by extended, non-variable
emission.  The latter method should be more reliable for measuring the
ratio of the core to knot A (assuming the aspect smearing affects both
components equally), but cannot be used for comparing countrates for a
given feature from different observations.

For the `adjoining rectangle' method, we chose in each map the same central
reference point lying on the line joining the two peaks, at about
the location of the saddle point in X-ray brightness between the core
and knot~A.  This reference point was derived from the
3$^{\prime\prime}$ smoothed contour plots.  Using a rotation of
20$^{\circ}$, two adjoining boxes of dimensions $\Delta\,x^{\prime} =
16^{\prime\prime}$, $\Delta\,y^{\prime} = 26^{\prime\prime}$ were
constructed. (The primed coordinates refer to the rotated frame.)
For the background, we joined the two measuring boxes to make the sum
($32^{\prime\prime}\times 26^{\prime\prime}$) and used a
$10^{\prime\prime}$ border around this box (all centered on the
reference point).  We also use this same background frame for the
r$=6^{\prime\prime}$ circular aperture.  A rough sketch of this
geometry is shown in Figure~2. 

For each map, positions of the core and knot A were determined by the
detection algorithm in IRAF/PROS and checked with contour diagrams of the
smoothed images (Fig.~1).  For the circular apertures, these positions
defined the centers of the circles.  The two Einstein observations were
reduced in the same manner as the ROSAT data.

\section{VARIABILITY OF THE CORE AND KNOT A}

\subsection{Ratio of core to knot A}

	The observable which is least affected by systematic
differences between EO and ROSAT, and between differences in quality
of aspect solutions, is the ratio of the flux of the core to knot A.
While there will always be some degree of `contamination' in the
measuring circles because the background is estimated within a region
somewhat removed from the core and knot A, this effect is minimized by
using the small area of the circular apertures.  We also
expect that whatever loss occurs from aspect smearing will affect both
core and knot A equally, so such an effect will only serve to reduce
any real changes in the ratio. 

The only important uncertainty which we have identified is the
difference in the effective areas of EO and ROSAT.  If the core were
significantly harder than knot A, then EO would find a different ratio
for the core to knot A than would ROSAT, supposing that they observed the
source at the same time.  The unknown spectral distributions of the
core and knot A lead to an uncertainty of roughly $\pm$25\% (for a
reasonable range of spectral distributions, see below) when comparing
EO with ROSAT countrates.  The results for the circular apertures and
ratios are given in Table~2 and plotted in Figure~3a.

\subsection{Countrates for the core and knot A}

While the HRI on Einstein was very similar to that on ROSAT, the
quantum efficiency and effective area were much smaller and the energy
band was wider.  Both the PIMMS software (a multi-mission tool
distributed by the High Energy Astrophysics Science Archive Research
Center, Goddard Space Flight Center) and the 'xflux' task in IRAF/PROS
use the appropriate effective areas of mirror/detector pairs, and
allow convolution with simple spectral shapes.  The conversion factor,
$C$, (ROSAT c/s $= C \times $EO c/s) for a power law spectrum with
energy index $\alpha=1.3$ $[S \propto \nu^{-\alpha}]$ and column
density, $\log$\,N$_{H}$ = 20.38 (the values used in BSH), is 1.75 for
PIMMS and 2.05 for xflux.  These numbers may be compared to conversion
factors deduced from Table 4 of the HRI Calibration Report (David et
al. 1995) for various supernova remnants where the conversion factors
are generally greater than 3.  In view of this uncertainty, we have
chosen to use the M87 cluster gas itself as our primary intensity
calibrator.  To do this, we measured the countrate in a circle of
radius 276$^{\prime\prime}$ centered on the reference point described
above.  For the background, we used an annulus with radii of
280$^{\prime\prime}$ and 300$^{\prime\prime}$.  We excluded from the
circle the inner box ($32^{\prime\prime}\times 26^{\prime\prime}$,
rotated by 20$^{\circ}$) which contains the core and knot A.  The
correction factors necessary to obtain the 95Jun value (which is taken
as the fiducial point) are listed in Table 3.  They may be compared
with results for bremsstrahlung spectra with kT = 2 keV and
$\log$\,N$_{H}$=20.38 of 2.1 for PIMMS and 2.45 for xflux.

As discussed above, we have based our countrate estimates on intensity
measurements in $16^{\prime\prime}\times 26^{\prime\prime}$ boxes.
The countrates for the core and knot A from both instruments are given
in Table 4 and plotted in Figure 3b.  The chief uncertainty is the
correction factor used to convert Einstein countrates to ROSAT values.
This factor is derived from the countrates of the cluster gas which is
believed to have a temperature close to 2 keV (Fabricant, Lecar \&
Gorenstein 1980; Nulsen \& B\"{o}hringer, 1995). Consequently, the
conversion factor could be as much as 35\% smaller if the spectral
distribution of the core or knot A were to be extremely different from
that of the cluster gas.  This spectral uncertainty precludes a
definitive statement about the history of the variability on the 10
year time scale covered by the Einstein and ROSAT observations.
However, the Einstein data alone show that the core intensity
increased by 16\% (4$\sigma$) between 79Dec and 80Jul whereas knot A
increased by less than 7\% during the same period (1.5$\sigma$).
During the 3.5 year ROSAT coverage, knot A declined by 16\% and the
core increased by 12\% (3$\sigma$) from 92Jun to 95Jun and declined by
17\% (7$\sigma$) in the following 6 months.

We also searched for short time-scale variability during each ROSAT
observation (one to three days of length).  Kolmogorov-Smirnov and
Cramer-von Mises one-sample goodness-of-fit tests were performed using
the circular apertures for both the core and knot A in order to test
the null hypothesis of constant source intensity.  The only instance
where the statistic exceeded 99\% was for knot A, 95Jun. The light
curve shows a 20\% enhancement for about 12 hours on 95Jun09.  This
behavior could be caused by aspect problems, and will be investigated
at a later time.

\section{DISCUSSION}

The magnitude of the characteristic changes is of order 0.02
count\,s$^{-1}$.  The conversion of ROSAT countrate to luminosity at
M87 (assumed to be 16~Mpc distant) varies between $3\times 10^{41}$
erg/count for soft spectral distributions (power laws with $\alpha =
2.5$ or bremsstrahlung spectra with kT = 0.2 keV) to $14\times
10^{41}$ erg/count for harder spectra ($\alpha = 0.2$ or kT = 10 keV).
Consequently, the changes we have observed are of order
$\Delta\,L_{x}$(0.5-3keV) = $10^{40}$ erg/s; substantially 
larger than the typical luminosities of galactic binaries (Tanaka \&
Lewin 1995; van Paradijs \& McClintock 1995).
  
Conventional explanations for the X-ray emission from the cores of
galaxies containing massive black holes are either thermal emission
from the putative accretion disk or non-thermal emission, possibly
associated with the inner jet, which may be strongly beamed.  Either
of these models can easily accommodate the observed variability and
short timescales.  Additionally, larger fractional changes have been
observed for other AGN (e.g.\ the Seyfert I galaxies reported by
Boller, Brandt \& Fink 1996).  Rapid variations have also been seen in
VLBI observations of the nucleus. Junor \& Biretta (1995) have found
evidence for changes in the jet structure in 1.3 cm VLBI images on
very small scales ($\approx$0.01\,pc) accompanied by a decrease in the
core brightness of $\approx30$\,\% over 5\,months in 1992.  In 1977, a
`flare' was observed with 2.3\,GHz VLBI; the amplitude changed by 30\%
over 4 months (Morabito, Preston \& Jauncey 1988).  While there are no
simultaneous flux measurements in the radio or optical bands, the
sporadic data which are available show the same sort of behavior as
that in Figure 3b.  The 2 cm radio core flux density (0.15''
resolution) increased by 13\% between 93Jan and 94May.  (These VLA
data are described in Biretta, Zhou \& Owen (1995).)  The ultraviolet
flux from the core (0.04'' resolution) decreased by a similar amount
between 94Aug and 95Jul (Biretta, Sparks \& Macchetto 1996).  These
data are consistent with a maximum in the core's lightcurve occurring
in mid 1994.

For knot A, the situation is different.  Even if the apparent decrease
of more than 10\% between 1980 and 1992 is uncertain because the
spectral distribution is unknown, the secular dimming of 16\% between
1992Jun and 1995Dec is a 3 $\sigma$ effect.  The observed decline (of
order 4\%/yr) is consistent with the halflife (12.8 yr) estimated by
BSH for relativistic electrons producing X-ray synchrotron emission in
a $200\mu$G field.  However, the physical size of knot A is known from
radio and optical data to be of order 70 parsec by a few parsecs
(i.e.\ a thin shock disk).  Therefore, the observed decrease could be
explained by (a) 100\% variability from a region of order a light year
across which, at maximum, contributes about 20\% of the knot A flux;
(b) the entire X-ray emitting volume of knot A could be substantially
smaller than the diameter of the disk which produces the optical and
radio emission; or (c) relativistic effects such as a change of a
beaming angle might be present.  For the former cases, the time scale
of the observed decrease favors synchrotron emission as the X-ray
emission process.  BSH estimate cooling times for thermal models of
over 100,000 years, and inverse Compton models always involve
relativistic electrons with substantially lower energies (and hence
longer lifetimes) than those required to produce X-ray synchrotron
emission.  Additional ROSAT observations have been approved
to monitor M87 at 6 month intervals and contemporary optical and radio
observations are planned.

\section*{ACKNOWLEDGMENTS}

Reduction of ROSAT data at SAO was supported by NASA contract
NAS5-30934 and NASA grant NAG5-2957 provided partial support of DEH
and JAB.  J.\ D.\ Silverman performed the Einstein measurements.  We
thank several of our colleagues at the CfA for assistance in
understanding the change in the EINSTEIN HRI sensitivity and the
referee, A. Edge, who made several useful suggestions for improving
the manuscript.

\section*{REFERENCES}

\ref{
Biretta J.A., Stern C.P., Harris D.E., 1991, AJ, 101, 1632}
\ref{
Biretta J.A., Zhou F., Owen F.N., 1995, ApJ, 447, 582}
\ref{
Biretta J.A., Sparks W.B., Macchetto F. 1996, in preparation}
\ref{
Boller Th., Brandt W.N., Fink H., 1996, A\&A, 305, 53}
\ref{
David L.P., Harnden Jr. F.R., Kearns K.E., Zombeck M.V., 1995,
"The ROSAT HRI Calibration Report", available from RSDC MS-3, SAO,
Cambridge, MA 02138, USA, and via WWW:
http://hea-www.harvard.edu/rosat/rsdc\_www/HRI\_CAL\_REPORT/hri.html}
\ref{
Fabricant D., Lecar M., Gorenstein P., 1980, ApJ, 241, 552}
\ref{
Junor W., Biretta J. A., 1995, AJ, 109, 500}
\ref{
Morabito D.D., Preston R.A., Jauncey D.L., 1988, AJ, 95, 1037}
\ref{
Mushotzky R.F., Done C., Pounds K.A., 1993, ARA\&A, 31, 717} 
\ref{
Nulsen P.E.J., B\"{o}hringer H., 1995, MNRAS, 274, 1093}
\ref{
Schreier E.J., Gorenstein P., Feigelson E.D., 1982, ApJ, 261, 42}
\ref{
Tanaka Y., Lewin W.H.G., 1995, in Lewin W.H.G., van Paradijs J., van
den Heuvel E. P. J., eds, X-ray Binaries, Cambridge Univ. Press,
Cambridge, p.126}
\ref{
van Paradijs J., McClintock J.E., 1995, in Lewin W.H.G., van Paradijs
J., van den Heuvel E. P. J., eds, X-ray Binaries, Cambridge
Univ. Press, Cambridge, p.58}

\newpage

\begin{center}
\begin{tabular}{llll} \\
\multicolumn{4}{c}{Table 1  Observations} \\
\\
Date		&Seq \#		&Livetime	&Comments \\
				 &&(secs) \\
\\
1979Jul05	&H282		&72662	&See BSH for details \\
1980Dec09	&H10316		&49249	&See BSH for details \\
1992Jun07	&wh700214	&13954	&good aspect \\
1995Jun09	&us701712	&44264	&poor aspect \\
1995Dec16	&us701713	&40362	&poor aspect
\end{tabular}
\end{center}
\smallskip

Notes: Deadtimes for Einstein are assumed to be 4\%; for ROSAT they are 2\%.

\bigskip
\bigskip

\begin{center}
\begin{tabular}{llllll} \\
\multicolumn{6}{c}{Table 2   Circular Aperture Counts for the Ratio
of Core / Knot A} \\
& \\

&1979&1980&1992Jun&1995Jun&1995Dec \\
& \\
core raw&2185&1535&1281&4465&3267 \\
core net&1842 (48)&1325 (40)&1084 (37)&3843 (68)&2702 (58) \\
& \\
knot A raw&2082&1383&1022&2892&2578 \\
knot A net&1742 (47)&1174 (38)&826 (33)&2271 (55)&2012 (52) \\
& \\
ratio (net)&1.06 (.04)&1.13 (.05)&1.31 (.07)&1.69 (.05)&1.34 (.05)
\end{tabular}
\end{center}
\smallskip

\noindent
Note: 1 $\sigma$ errors are given in parentheses; those on the ratios are
the sums of the errors on each component, taken in quadrature.  The
Einstein ratios may be compared to that derived in the BSH paper
where extensive image processing was performed so that background
contamination was minimized: (Core+knot D)/(knot A+knot B) = 1.15

\newpage

\begin{center}
\begin{tabular}{lrrlcrl} \\
\multicolumn{7}{c}{Table 3	Intensity Calibration Based on
Extended Thermal Emission} \\
& \\
Date&Raw&\multicolumn{2}{c}{Net countrate}&\hspace*{.5in}&\multicolumn{2}{l}{CorFac
to 95Jun} \\
&(cnts)&\multicolumn{2}{c}{(c/s)} \\
& \\
79Jul		&124324	&0.780 &(0.010)	&&2.940 &(1.3\%) \\
80Dec		&72388	&0.671 &(0.012)	&&3.418 &(1.8\%) \\
92Jun		&57660	&2.183 &(0.034)&&1.050 &(1.6\%) \\
95Jun	&185689	&2.293 &(0.019)	&&1.000 &(0.8\%) \\
95Dec	&167682	&2.304 &(0.020)	&&0.995 &(0.9\%)
\end{tabular}
\end{center}
\smallskip

\noindent
Note: Raw counts are the value for the r$=276^{\prime\prime}$ circle
centered on the reference point, minus the counts in the rotated box
($32^{\prime\prime}\times 26^{\prime\prime}$).  The net countrate is
based on the background subtraction of the 280$^{\prime\prime}$ to
300$^{\prime\prime}$ annulus.  The 5\% difference between 92Jun and
95Jun is ascribed to the change in high voltage (94Jun; see the HRI
Calibration Report, David et al.\ 1995).  The 16\% difference between
the two Einstein observations is reasonably close to the 12.3\% drop
in sensitivity expected in the 17 months between the two observations.
This secular change in the sensitivity is scaled from the estimate of
8.7\%/yr derived from observations of a number of supernova remnants
and Abell 496 (Seward and Martenis, internal Einstein Memo of 1988 Jul
21).  1 $\sigma$ errors are given in parentheses.

\bigskip
\bigskip

\begin{center}
\begin{tabular}{lcllcll} \\
\multicolumn{7}{c}{Table 4    Core and Knot A Countrates} \\
& \\
&\hspace*{.5in}&\multicolumn{2}{c}{CORE}&\hspace*{.5in}&\multicolumn{2}{c}{KNOT
A} \\
Date&&Box&Net countrate&&Box&Net countrate \\
&&(cnts)&(c/ksec)&&(cnts)&(c/ksec) \\
& \\
79Jul&&4133&116.5 (2.8)&&3680&98.0 (2.6) \\
80Dec&&2722&135.5 (3.8)&&2275&104.5 (3.5) \\
92Jun&&2336&123.7 (3.9)&&1873&88.4 (3.5) \\
95Jun&&8417&138.6 (2.2)&&5782&79.0 (1.8) \\
95Dec&&6747&115.1 (2.1)&&5104&74.5 (1.9)
\end{tabular}
\end{center}

\smallskip

\noindent
Note: the box counts (columns 2 and 4) are given without any
corrections but the countrates (columns 3 and 5) are corrected for
background in the 10$^{\prime\prime}$ wide frame and have been
multiplied by the appropriate correction factor from Table~3.  1 $\sigma$
errors are given in parentheses.

\newpage

Figure~1: Contour diagrams of the data with a 3$^{\prime\prime}$ FWHM
Gaussian smoothing function.  The maps have been scaled by
$10^{6}$/livetime to change the units to counts/pixel/Megasec.  The
pixel size is 0.5$^{\prime\prime}$ and the contour levels are
logarithmic: 40, 53, 70, 93, 124, 164, 218, and 290 c/pix/Ms.  (a)
1992Jun; (b) 1995Jun; (c) 1995Dec.

\vspace*{1in}

Figure~2: A grey scale image of M87 with the approximate geometry for
intensity measurements shown. The inner rectangle is divided into 2
equal areas for the `adjoining rectangle' method discussed in the
text. 

\vspace*{1in}

Figure~3: 	Variability Results\newline

For calendar dates, see Table~1.  (a) the ratio of net counts in
r$=6^{\prime\prime}$ circular apertures centered on the core and knot
A.  (b) the countrates (c/ksec) for the box measurements of the core
(circles) and knot A (squares).  The correction factors used are those
from Table~3.  Included as a control (the x$^{\prime}$s) are the
differences between the countrates in the background frame and the
countrates in a circle (r$=12^{\prime\prime}$) located
45$^{\prime\prime}$ to the SE of the reference point (a region where
the X-ray surface brightness is without large spatial gradients and is
approximately 60\% of the average frame value).

\bigskip

For the Einstein data, additional uncertainties caused by the unknown
spectral distribution of components, are roughly $+6\%$ (for harder
spectra, up to 10 keV, or $\alpha$ down to 0.3) and $-35\%$ (for
softer spectra, down to 0.3 keV or $\alpha$ up to 2.4).  Similar
uncertainties would apply to the ROSAT data only if the variability
was accompanied by a significant change in spectral distribution.

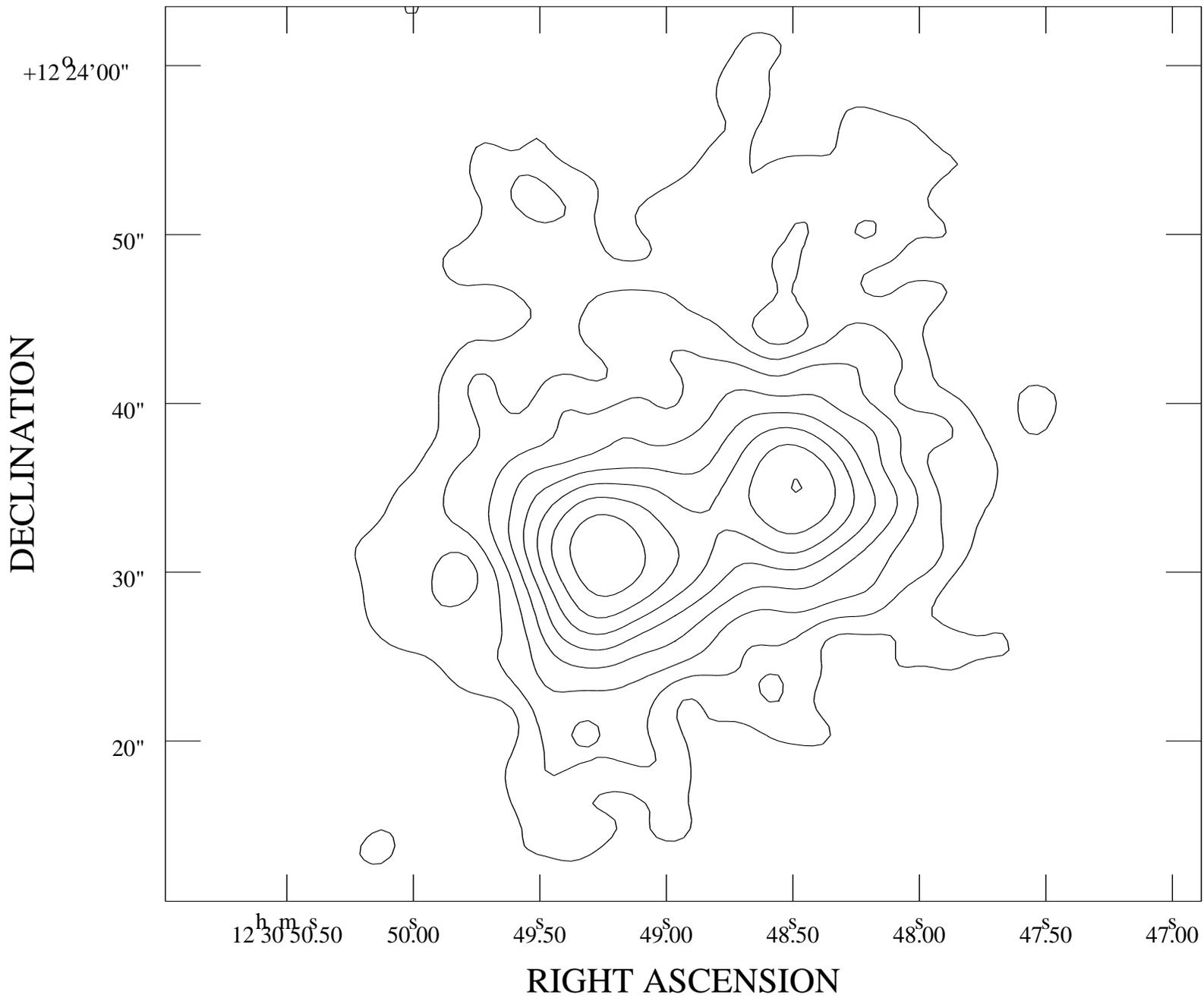
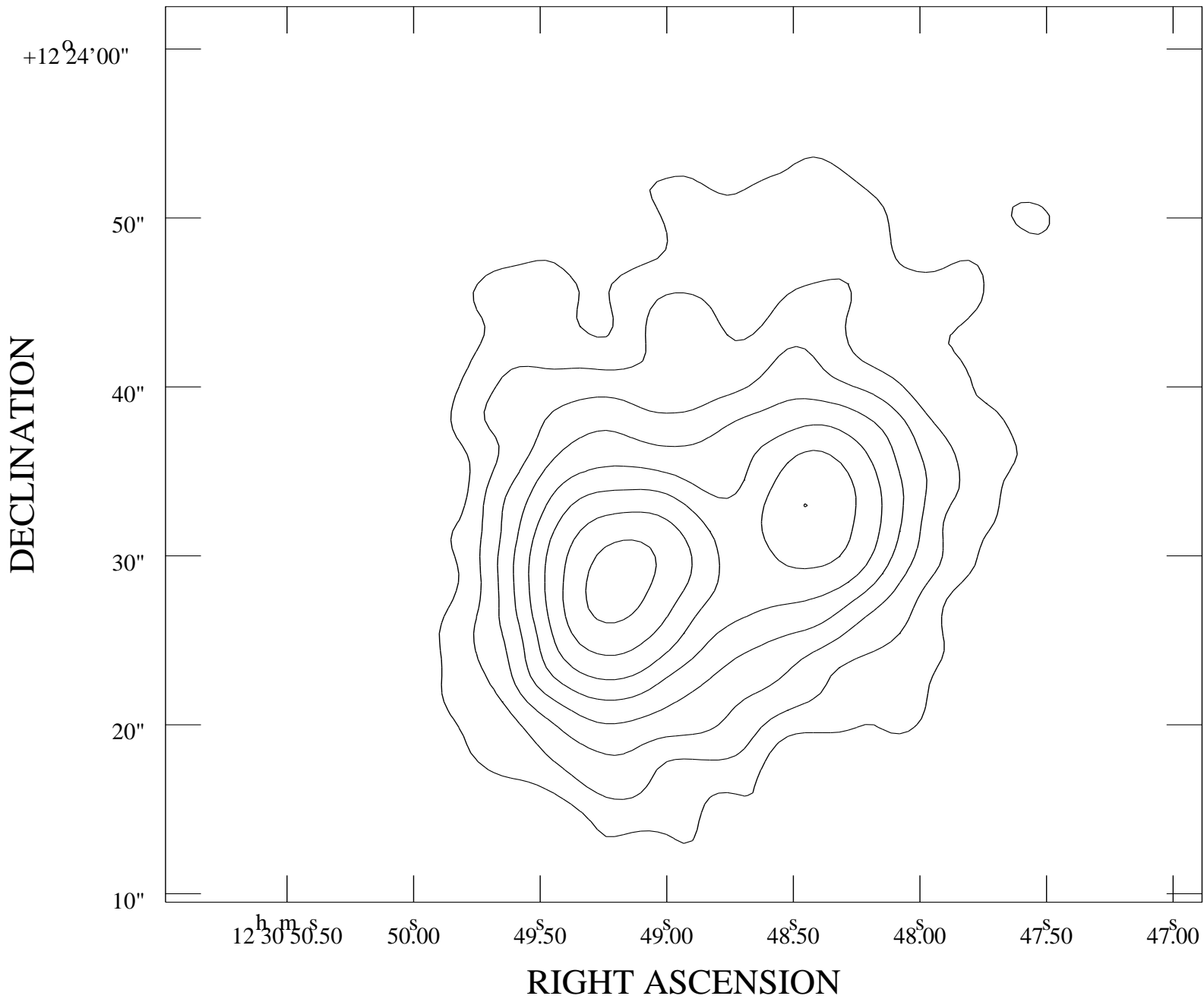
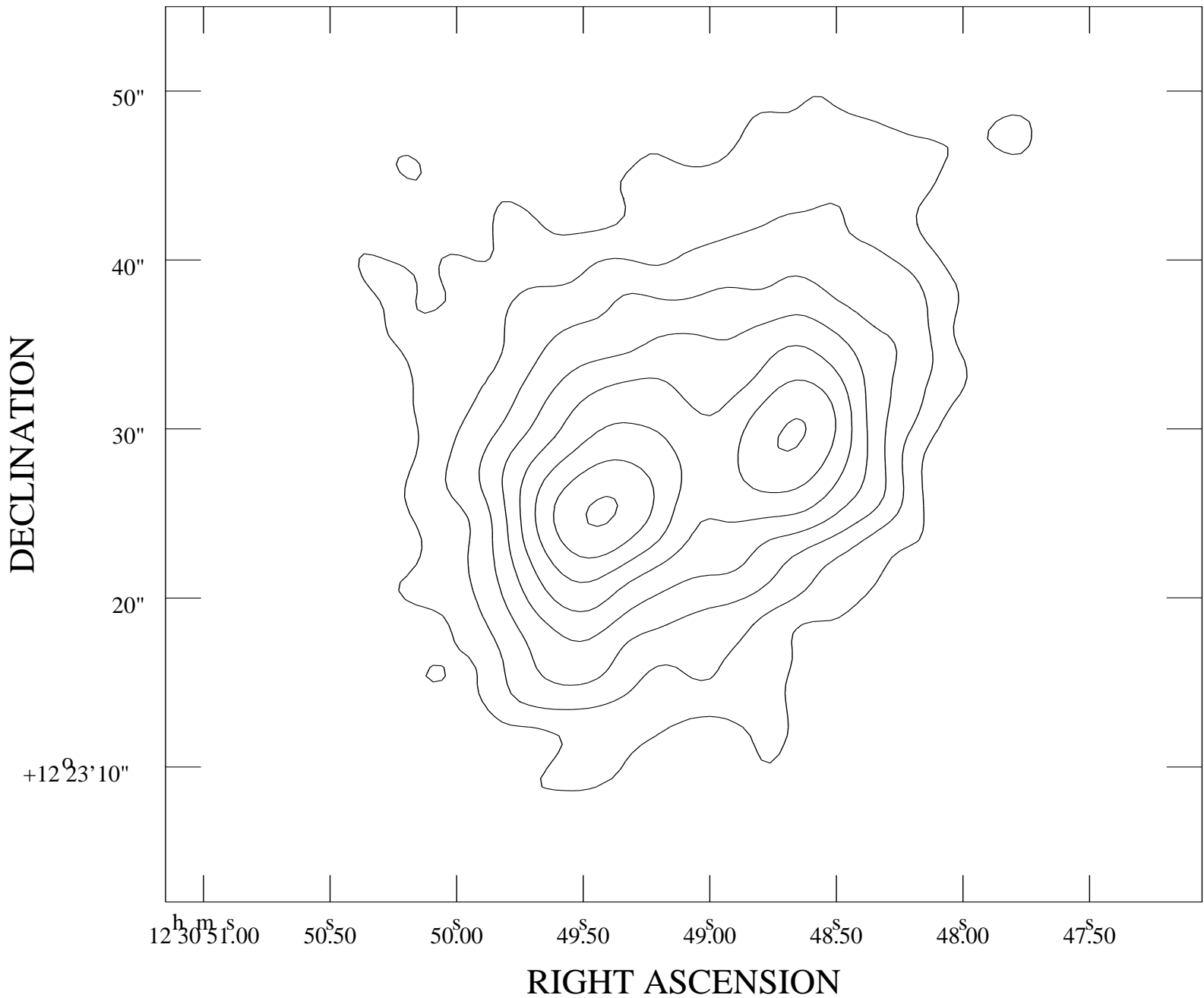
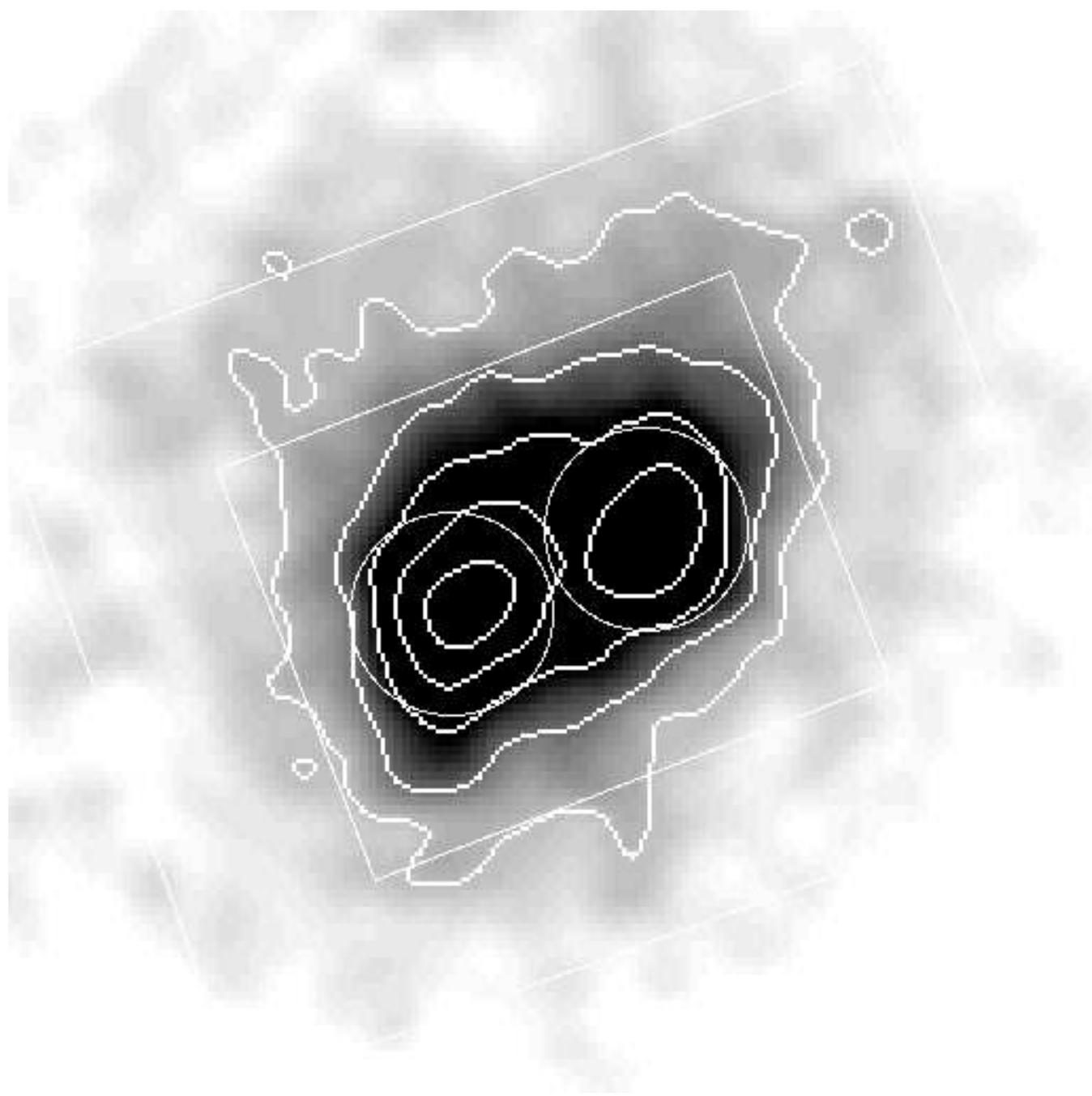
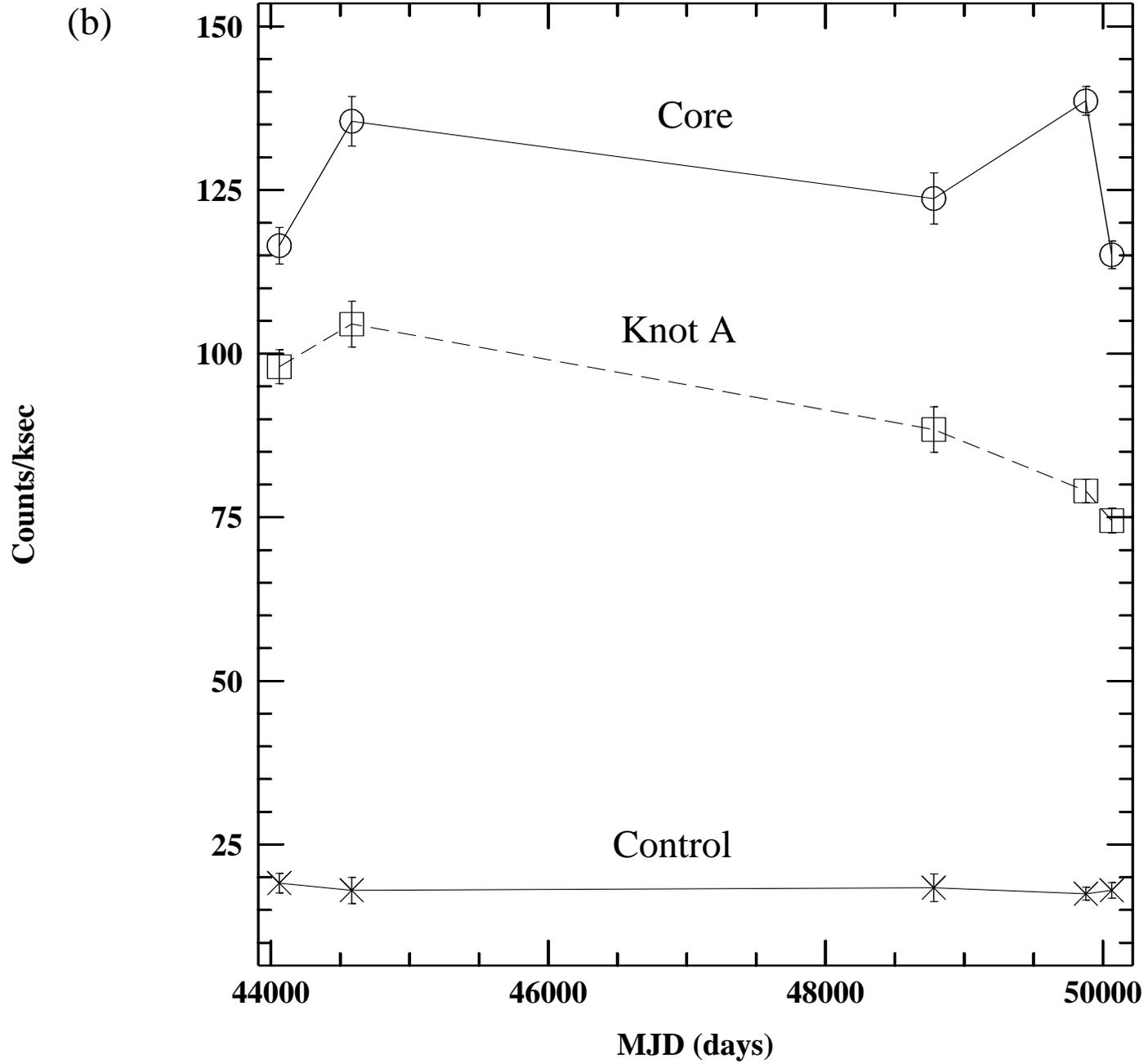
\end{document}